\documentclass[journal]{IEEEtran}

\usepackage{graphicx}  
\usepackage{amssymb}
\begin{document}
%
\title{Improved modeling of Coulomb effects in nanoscale Schottky-barrier FETs}
%
%
\author{Klaus~Michael~Indlekofer,
        Joachim~Knoch,
        and~Joerg Appenzeller
\thanks{K. M. Indlekofer and J. Knoch are with
the Institute of Bio- and Nanosystems, IBN-1 and the cni - Center
of Nanoelectronic Systems for Information Technology, Research
Centre J\"ulich GmbH, D-52425 J\"ulich, Germany
(e-mail: m.indlekofer@fz-juelich.de).}
\thanks{J. Appenzeller is with IBM T. J. Watson Research Center, Yorktown Heights, NY 10598, USA.}}
%
%
%
\markboth{IEEE Transactions on Electron Devices,~Vol.~1, No.~1,~November~2006}{Indlekofer \MakeLowercase{\textit{et al.}}: Coulomb effects in nanoscale SB-FETs}
%



\maketitle

\begin{abstract}
We employ a novel multi-configurational self-consistent Green's
function approach (MCSCG) for the simulation of nanoscale
Schottky-barrier field-effect transistors. This approach allows to
calculate the electronic transport with a seamless transition from
the single-electron regime to room temperature field-effect
transistor operation. The particular improvement of the MCSCG stems
from a division of the channel system into a small subsystem of
resonantly trapped states for which a many-body Fock space becomes
feasible and a strongly coupled rest which can be treated adequately
on a conventional mean-field level. The Fock space description
allows for the calculation of few-electron Coulomb charging effects
beyond mean-field.

We compare a conventional Hartree non-equilibrium Green's function
calculation with the results of the MCSCG approach. Using the
MCSCG method Coulomb blockade effects are demonstrated at low
temperatures while under strong nonequilibrium and room
temperature conditions the Hartree approximation is retained.
\end{abstract}


%
\IEEEpeerreviewmaketitle

\section{Introduction}
\PARstart{O}{ne} of the major challenges for the simulation of
nanoscale field-effect transistors (FET) consists in an adequate
description of the Coulomb interaction within the transistor
channel: a proper simulation approach has to account for the
Coulomb interaction of a few fluctuating electrons and at the same
time has to be able to describe non-equilibrium transport in an
open nanosystem.

For the correct many-body description of the Coulomb interaction
with the inclusion of contact coupling and nonequilibrium injection
conditions, Fock space approaches such as realtime renormalization
group \cite{rtrg,rtrg2} or the Fock space Green's function
\cite{fsgf} are available. This class of methods provides kinetic
equations in Fock space, taking renormalization and dissipation due
to the coupling to the contacts (at least to some extent) and the
Coulomb interaction into account. Since these approaches involve the
$2^N$-dimensional Fock space for the considered single-particle
basis of $N$ states, they typically are restricted to $N\lesssim 10$
for practical reasons. In the limit of small coupling, the Fock
space description of the system can be approximated by a reduction
to rate equations \cite{been,gurv}, which deal with many-body
eigenstates of the uncoupled Hamiltonian obtained via exact
diagonalization \cite{ediag,pfann,indl00,indl}, containing a full
description of the Coulomb interaction.

Realistic modeling of a 1D semiconductor nanotransistor typically
involves a number $N$ of single-particle states (orbital or sites)
of up to a few hundred, rendering a full numerical Fock space
description impossible due to the exponential scaling of the
resulting many-body space dimensions. Furthermore, most of the
potentially current-carrying single-particle states are stronlgy
coupled to the contacts, and thus, the picture of a weakly coupled
system in general becomes inadequate. The nonequilibrium Green's
function (NEGF) approach \cite{datta,lake,guo,pul} in a mean-field
approximation provides reasonble scalability, however, in principle
lacks the description of few-electron Coulomb charging effects which
become apparent in the case of resonantly trapped states in
particular at lower temperatures. A possible solution is the
combination of the numerically well scaling mean-field NEGF with a
Fock space description for those states where many-body Coulomb
effects may become important for the device characteristics. In this
context, we have recently proposed a multi-configurational
selfconsistent Green's function approach (MCSCG) \cite{indl,indl2}
for the realistic simulation of nanodevice systems under
application-relevant conditions with reasonable numerical efforts.

In the following sections, we will outline the main ideas and the
algorithm behind the MCSCG and demonstrate its strengths by
comparing a conventional Hartree NEGF calculation with the results
of the MCSCG approach, providing significantly extended information
to our recent conference contribution \cite{indl3}. Using the MCSCG,
Coulomb blockade effects are demonstrated at low temperatures while
under strong nonequilibrium and room temperature conditions the
Hartree approximation is retained.

\section{Multi-configurational approach}
\begin{figure}
\begin{center}
\includegraphics[width=3.5in]{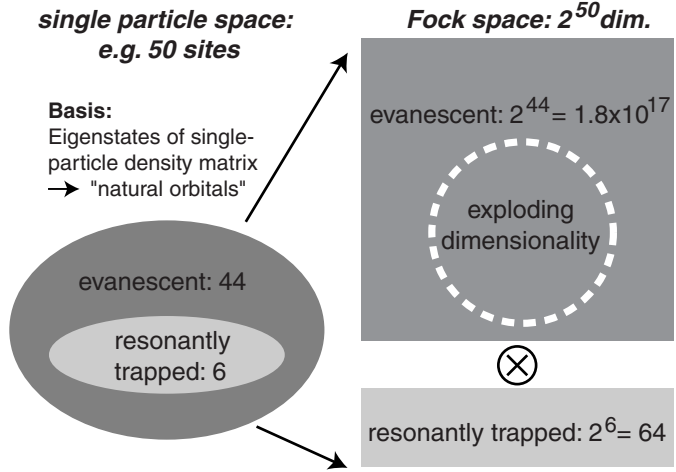}
\caption{Example for the division of the single-particle space into
two subspaces. Only for resonantly trapped states (6 in the
example), a Fock space description is employed. The rest
(evanescent) is treated by a mean-field description.}
\label{fig_subspace}
\end{center}
\end{figure}

In order to handle systems with a large number $N$ of
single-particle states, the main idea of the MCSCG is to divide the
system into two subsystems: Resonantly trapped (i.e. weakly coupled)
states and those states that couple strongly to the contacts as
depicted in Fig.~\ref{fig_subspace}. Within the subspace of
$N'\lesssim 10$ resonantly trapped states, which require a many-body
description of the Coulomb interaction, a Fock space method will be
applied, whereas the rest ($N-N'$) is treated adequately on an
approximated NEGF level. The eigenstates $\{|\kappa\rangle\}$ of the
many-body statistical operator (or Hamiltonian, depending on the
Fock space method) within the resonantly trapped subspace will be
referred to as configurations with weights $\{w_{\kappa}\}$
corresponding to the respective eigenvalues. Thus, the
configurations and their weights follow from a Fock space
calculation, taking the detailed Coulomb interaction within this
subspace into account. The many-body statistical operator $\rho$ of
the considered system can be written in the general form
\begin{equation}
\rho=\sum_{\kappa}w_{\kappa}P_{\kappa}\otimes\rho_{rest}[\kappa],
\end{equation}
where $P_\kappa\equiv|\kappa\rangle\langle\kappa|$ denotes the
projection operator corresponding to the eigenstate
$|\kappa\rangle$, and $\rho_{rest}$ is the many-body statistical
operator of the rest, which may depend on the configuration
$\kappa$. Motivated by this form, we define a configuration-averaged
Green's function $\bar{G}$ of the system as
\begin{equation}
\bar{G}=\sum_{\kappa}w_{\kappa}G[\kappa],
\end{equation}
where $G[\kappa]$ corresponds to
$P_{\kappa}\otimes\rho_{rest}[\kappa]$ and shall fulfill Dyson's
equation with a suitable contact coupling plus Coulomb selfenergy
approximation $\Sigma[\kappa]$. In the simplest case,
$\Sigma[\kappa]$ may be of a decoupled mean-field form \cite{indl},
which is adequate for temperatures well above the corresponding
Kondo temperature.

As for the Fock subspace of the resonantly trapped states, the
projected many-body Hamiltonian contains the reduced single-particle
and Coulomb terms. Furthermore, coupling to the contacts with
nonequilibrium carrier injection and coupling (tunneling) to the
rest of the system is described by means of selfenergy kernels,
depending on the chosen Fock space method. For the latter, various
choices are possible, for example: Exact diagonaliztion with Dyson's
equation as subsidiary condition \cite{indl,indl2}, real-time
renormalization group (RTRG) \cite{rtrg,rtrg2} or Fock space Green's
functions \cite{fsgf}. In the following, we will discuss the first
option, based on exact diagonalization. In this case, the many-body
statistical operator is assumed to be diagonal in the eigenbasis of
the resonant subspace Hamiltonian. Here, the eigenvalues
$w_{\kappa}$ are determined such that the resulting many-body
Green's function $G_{MB}$ in Lehmann representation fulfills Dyson's
equation within the resonant subspace. In the simplest
implementation, $\{w_{\kappa}\}$ is chosen such that the spectral
peaks of $G^<_{MB}$ match those of $\bar{G}^<$ \cite{indl}.
\begin{figure}
\centering
\includegraphics[width=3.5in]{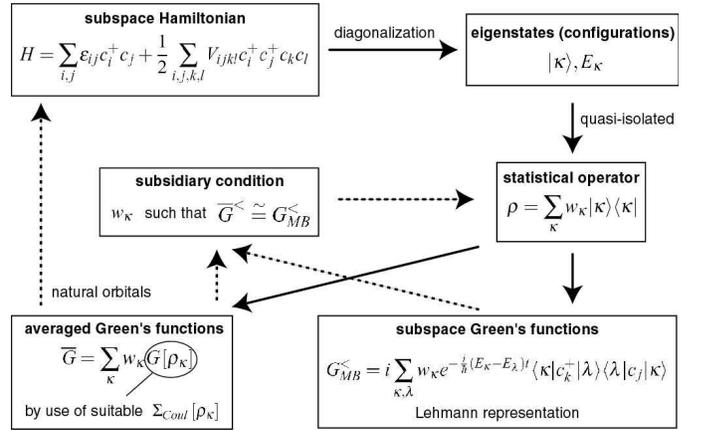}
\caption{Flowchart of the MCSCG algorithm. In the shown case, exact
diagonalization is employed as the simplest Fock subspace method.
Arrows visualize the flow of the selfconsistency loop.}
\label{fig_flowchart}
\end{figure}

As an overall selfconsistency condition, the resonantly trapped
states experience a mean-field interaction of the rest, whereas the
rest is subject to the set $\{\Sigma[\kappa]\}$ of selfenergies
originating from the resonant many-body configurations $\kappa$ and
its own mean-field interaction. Finally, for the identification of
resonantly trapped states, the single-particle eigenstates of the
single-particle density-matrix are employed (so-called natural
orbitals). The latter follows directly from the Green's function
$\bar{G}^<$ as part of the multi-configurational selfconsistency
procedure. (Note that each individual $G[\kappa]$ need not be
selfconsistent with its respective $\Sigma[\kappa]$.) In turn,
resonantly trapped states are defined as those single-particle
eigenstates that exhibit a level broadening (determined from the
selfenergies) below a given threshold. Fig.~\ref{fig_flowchart}
illustrates the details of the algorithm as a flowchart.

For the calculation of expectation values of single-particle
observables (e.g., electron density, current, spin density, etc.),
the selfconsistent Green's function $\bar{G}^<$ is employed as an
approximation for the unknown exact $G^<$. Optionally, within the
resonant subspace, one can use the many-body result (such as
$G_{MB}$, the reduced statistical operator, dissipation kernel,
etc.) for the evaluation of arbitrary expectation values, in
particular contour-ordered correlation functions of arbitrary order.

\section{Simulation of Nanoscale Schottky-barrier FETs}
\begin{figure}
\centering
\includegraphics[width=2.5in]{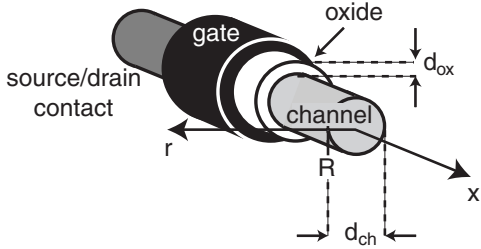}
\caption{Schematic sketch of a nanowire MOSFET with coaxial gate.
} \label{fig_schema}
\end{figure}

In order to demonstrate the strengths of the MCSCG approach, we
consider a one-dimensional (1D) coaxially gated nanowire transistor
with Schottky-barrier source and drain contacts since deviations
from a mean-field approximation become most apparent in a system
with quasi-bound states. Fig.~\ref{fig_schema} shows a schematics of
such a nanowire transistor where we assume a channel length of
$L=20$nm, a diameter of $d_{\rm ch}=4$nm and a gate oxide thickness
of $d_{\rm ox}=10$nm. (Since the spin degree of freedom is included
and a site spacing of $a=1$nm is employed, we thus consider $N=40$
single-particle states.) Such a nanowire transistor can in principle
be realized with a semiconducting InGaAs nanowire with $SiO_2$ as
gate dielectric. It has been shown that the electrostatics of
coaxially gated nanowire transistors can be well described by a
modified one-dimensional Poisson equation \cite{auth}. This Poisson
equation allows to easily calculate the Coulomb Green's function
which in turn enables the description of the classical
electrostatics and the screened interaction between electrons on
equal footing \cite{indl}.
For the following simulation results, the simplest MCSCG variant has
been employed, based on exact diagonalization (with $N'=6$
resonantly trapped states yielding 64 Fock space dimensions) with a
decoupled static selfenergy form.

\begin{figure}
\centering
\includegraphics[width=2.8in]{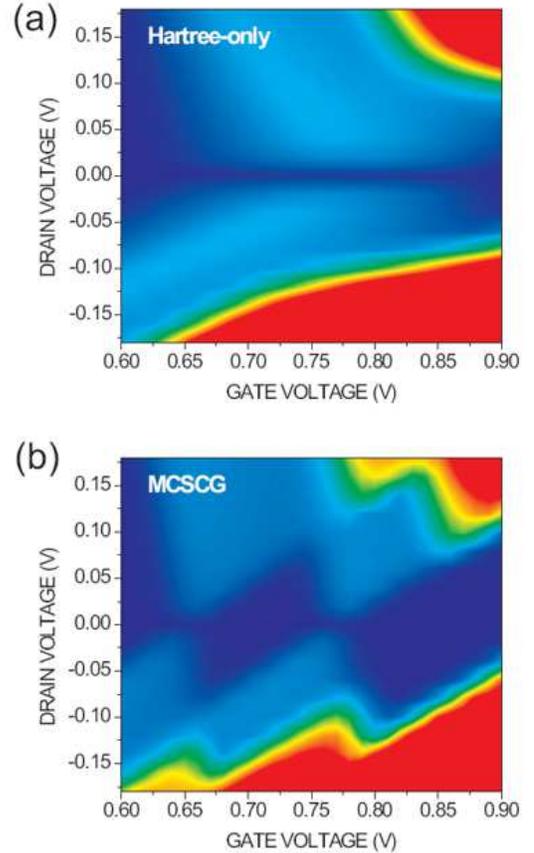}
\caption{(a) Color plot of the total drain current for the
Hartree-only case at $T=77$K (blue = 0nA, red = 5nA).  (b) Color
plot of the total drain current for the MCSCG case at $T=77$K. (blue
= 0nA, red = 5nA).} \label{fig_Diamonds77K}
\end{figure}
Fig.~\ref{fig_Diamonds77K} visualizes the simulated drain current
$I_{\rm D}$ for the single-electron transport regime ($T = 77$K) as
a color plot. In contrast to the Hartree-only calculation
(Fig.~\ref{fig_Diamonds77K}(a)), the MCSCG approach
(Fig.~\ref{fig_Diamonds77K}(b)) correctly reveals diamond-like
shaped patterns due to the quantized Coulomb interaction (as
predicted by the orthodox theory and observed in experiments). While
the MCSCG treatment is able to cope with the mixture of many-body
configurations, the Hartree theory only provides a mean interaction
potential for the description of the Coulomb interaction.

\begin{figure}
\centering
\includegraphics[width=3.5in]{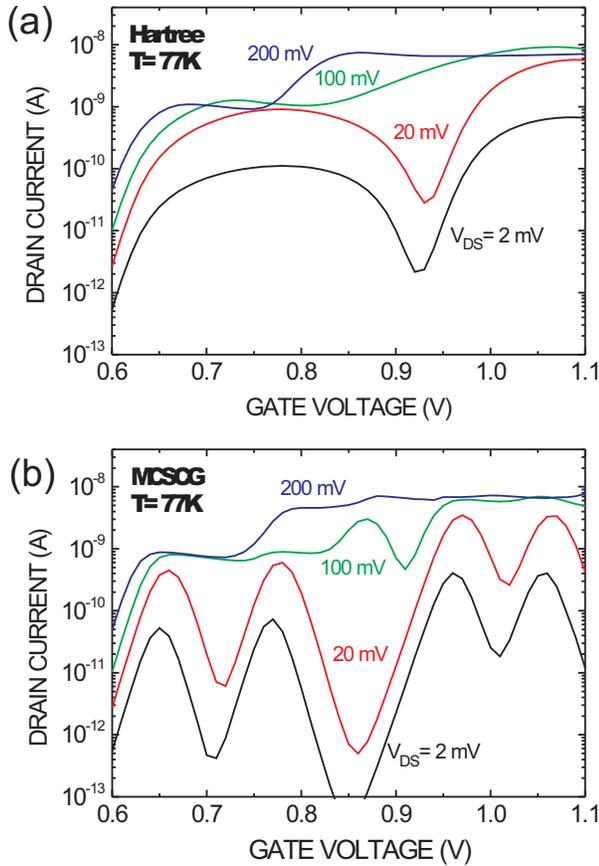}
\caption{(a) Transfer characteristics, Hartree-only at $T=77$K. (b)
Transfer characteristics, MCSCG approach at $T=77$K.}
\label{fig_Output77K}
\end{figure}
In addition, Fig.~\ref{fig_Output77K} shows $I_{\rm D}(V_{\rm GS})$
curves for different drain voltages $V_{\rm DS}$. In the MCSCG case
(Fig.~\ref{fig_Output77K} (b)), single-electron transport can be
identified in terms of Coulomb oscillations for the two lowest
$V_{\rm DS}$, whereas the Hartree-only simulation
(Fig.~\ref{fig_Output77K}(a)) lacks these features; the Hartree-only
case exhibits broader peaks solely due to the single-particle levels
of the system. However, with increasing $V_{\rm DS}$, both
approaches become equivalent. (Note that the sub-threshold regime
shows the regular behavior and has been omitted here.)

\begin{figure}
\centering
\includegraphics[width=3.5in]{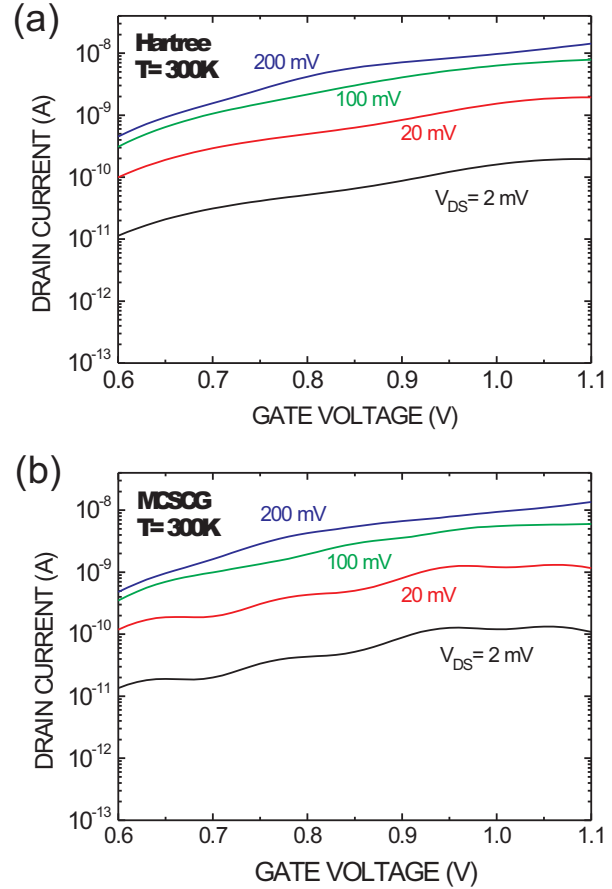}
\caption{(a) Transfer characteristics, Hartree-only at $T=300$K. (b)
Transfer characteristics, MCSCG approach at $T=300$K.}
\label{fig_Output300K}
\end{figure}
Finally, Fig.~\ref{fig_Output300K} shows the room temperature
($T=300$K) characteristics. Apart from the slight modulation in the
MCSCG calculation (Fig.~\ref{fig_Output300K}(b)), which is a remnant
of the Coulomb oscillation, the Hartree
(Fig.~\ref{fig_Output300K}(a)) and MCSCG (Fig.~\ref{fig_Output77K}
(b)) results are in good agreement. Effects beyond a mean-field
picture of the system will obviously have a significant impact on
application-relevant device properties such as the system
capacitance and the transconductance.

\section{Conclusion}
In summary, we have compared for the first time the conventional
Hartree NEGF with the MCSCG and have shown that the
multi-configurational approach is able to describe single-electron
charging effects in the low temperature limit for a realistic FET
structure. In case of strong nonequilibrium (with an almost depleted
channel) and room temperature conditions, the MCSCG and the
well-established Hartree approximation lead to equivalent results
for the discussed example of a nanowire MOSFET. As such, the MCSCG
yields a seamless transition from the single-electron transport
regime to transistor operation at room temperature. For realistic
FETs with a large number of sites where a full Fock space
formulation becomes impossible, the MCSCG permits a selfconsistent
Fock space treatment of states which are responsible for
few-electron charging effects.

\begin{biography}{Klaus Michael Indlekofer}
Klaus Michael Indlekofer received the Dipl.-Phys. degree in 1996
(summa cum laude) and the Dr. rer.nat. degree in 1999 (summa cum
laude), both in physics, from the RWTH Aachen (Germany). These works
were carried out in the Research Centre Juelich (Germany) and
focused on experimental as well as theoretical aspects of
single-electron effects in vertical III/V resonant tunneling
transistors. From 2000-2001 he was involved in the development of
the "WinGreen" transport simulation tool. In 2001-2002, he worked on
transport models for open quantum dot systems in the Department of
Electrical Engineering at the Arizona State University as a
Feodor-Lynen fellow (A. v. Humboldt foundation). In 2002 he returned
to the Research Centre Juelich and is involved in the development of
many-body real-time models for Coulomb effects in nanodevices.
\end{biography}

\begin{biography}{Joachim Knoch}
J. Knoch received the M.S. and Ph.D. degrees in physics from the
Technical University of Aachen, Aachen, Germany in 1998 and 2001,
respectively. At the University of Aachen he investigated quantum
transport in superconductor/semiconductor hybrids based on III-V
heterostructures as well as worked on the modeling and realization
of ultrashort-channel silicon MOSFETs. From September 2001 to
December 2002 he was with the Microsystems Technology Laboratory,
Massachusetts Institute of Technology, Cambridge, where he worked
on InP-HEMT devices. Currently, he is a Research Scientist at the
Institute of Thin Films and Interfaces, Research Center Juelich,
Germany. He is involved in the exploration of electronic transport
in alternative field-effect transistor devices such as CN
field-effect transistors, and ultrathin body Schottky-barrier
devices and MOSFETs based upon strained silicon.
\end{biography}


\begin{biography}{Joerg Appenzeller}
J. Appenzeller received the M.S. and Ph.D. degrees in physics from
the Technical University of Aachen, Germany in 1991 and 1995. His
Ph.D. dissertation investigated quantum transport phenomena in low
dimensional systems based on III/V heterostructures. He worked for
one year as a Research Scientist in the Research Center in
Juelich, Germany before he became an Assistant Professor with the
Technical University of Aachen in 1996. During his professorship
he explored mesoscopic electron transport in different materials
including carbon nanotubes and
superconductor/semiconductor-hybride devices. From 1998 to 1999,
he was with the Massachusetts Institute of Technology, Cambridge,
as a Visiting Scientists, exploring the ultimate scaling limits of
silicon MOSFET devices. Since 2001, he has been with the IBM T.J.
Watson Research Center, Yorktown, NY, as a Research Staff Member
mainly involved in the investigation of the potential of carbon
nanotubes for a future nanoelectronics.
\end{biography}




\end{document}